\documentclass[conference]{IEEEtran}
\usepackage{cite}
\usepackage{amsmath,amssymb,amsfonts}
\usepackage{algorithmic}
\usepackage{graphicx}
\usepackage{textcomp}
\usepackage{xcolor}

\makeatletter
\newcommand{\linebreakand}{%
  \end{@IEEEauthorhalign}
  \hfill\mbox{}\par
  \mbox{}\hfill\begin{@IEEEauthorhalign}
}
\makeatother
\usepackage{subcaption}
\usepackage{booktabs}
\usepackage{graphicx}
\usepackage{array}
\usepackage{adjustbox}
\usepackage{caption}
\usepackage{physics}
\usepackage{pifont}
\usepackage{xcolor}
\usepackage{multirow}
\newcommand{\yes}{\textcolor{green!55!black}{\ding{51}}}
\newcommand{\no}{\textcolor{red!70!black}{\ding{55}}}
\usepackage{url}

\def\BibTeX{{\rm B\kern-.05em{\sc i\kern-.025em b}\kern-.08em
    T\kern-.1667em\lower.7ex\hbox{E}\kern-.125emX}}
\begin{document}

\title{Practical Feasibility of Gradient Inversion Attacks in Federated Learning
\thanks{This work was funded by Vinnova (Sweden's Innovation Agency) through the LeakPro project, grant no.~2025-03066.}
}

\author{\IEEEauthorblockN{X}}

\author{\IEEEauthorblockN{Viktor Valadi}
\IEEEauthorblockA{\textit{Scaleout Systems}\\
Sweden}
\and
\IEEEauthorblockN{Lucas Beerens}
\IEEEauthorblockA{\textit{Scaleout Systems}\\
Sweden}
\and
\IEEEauthorblockN{Mattias Åkesson}
\IEEEauthorblockA{\textit{Scaleout Systems}\\
Sweden}
\and
\IEEEauthorblockN{Johan Östman}
\IEEEauthorblockA{\textit{Recorded Future}\\
Sweden}
\and
\IEEEauthorblockN{Fazeleh Hoseini}
\IEEEauthorblockA{\textit{AI Sweden}\\
Sweden}
\linebreakand
\IEEEauthorblockN{Salman Toor}
\IEEEauthorblockA{\textit{Scaleout Systems} \\
\textit{Uppsala University}\\
Sweden}
\and
\IEEEauthorblockN{Andreas Hellander}
\IEEEauthorblockA{\textit{Scaleout Systems} \\
\textit{Uppsala University}\\
Sweden}
}

\maketitle

\begin{abstract}
    Gradient inversion attacks are widely cited as a privacy threat in federated learning, yet a growing body of work suggests their effectiveness is narrower than early demonstrations implied. We present a feasibility analysis of optimization-based gradient inversion under an honest-but-curious server, investigating when, why, and to what degree it remains feasible as architectures, data, and training procedures approach realistic deployment. As a new diagnostic, we introduce the target-interpolation feasibility probe, which evaluates the attack objective along the path from an uninformed initialization to the target rather than locally around it, and tests at every step whether the direction toward the target is preferred over random perturbations of the same size. At a fraction of the cost of a full attack, it helps distinguish genuine information insufficiency from optimization failure and yields a graded signal of vulnerability rather than a binary verdict. To ensure our negative results reflect genuine resistance rather than a weak attack, we develop a stronger training-mode construction that, under the same assumptions as prior methods, yields more faithful reconstructions by recovering the client's exact normalization statistics to regularize against. Using these tools, we present, to our knowledge, the first evaluation of gradient inversion on object detection pipelines. All three detectors we evaluate resist reconstruction under training-mode updates; only YOLOv8-nano becomes vulnerable, and only when clients train in inference mode. Finally, an evaluation across modern ImageNet architectures shows resistance increasing when architectural and system complexity increase towards levels expected in deployment-grade systems. Taken together, our results indicate that, for image classification and detection under an honest-but-curious server, image reconstruction via optimization-based gradient inversion appears improbable in modern, performance-optimized configurations, even at batch size one.
\end{abstract}

\begin{IEEEkeywords}
federated learning, gradient inversion attacks, privacy attacks, training data reconstruction, deep neural networks, computer vision, object detection
\end{IEEEkeywords}

\section{Introduction}
Federated learning (FL)~\cite{mcmahan2017communication}, is a decentralized machine learning paradigm in which clients collaboratively train a shared model without exchanging their raw data. 
By keeping data local and transmitting only model updates, FL offers a compelling solution to data governance and privacy concerns, making it especially attractive in domains such as healthcare and finance.

However, this decentralized setup introduces new technical and security challenges. 
As outlined by~\cite{kairouz2021advances}, the three primary concerns in FL are robustness, communication efficiency, and privacy. A key aspect of robustness pertains to the system's resilience against malicious participants; for example, recent work has demonstrated that even a small fraction of adversarial clients can poison the global model \cite{sun2019can, valadi2023fedval, xhemrishi2025fedgt}. 
Communication efficiency remains a major bottleneck, particularly in cross-device FL scenarios, where uplink bandwidth is limited and client participation is sporadic \cite{konevcny2016federated, sattler2019robust, ekmefjord2022scalable}.
Finally, the gradients or model updates exchanged during training can leak sensitive information, potentially violating privacy. 
A growing body of research has demonstrated reconstruction of client data from model gradients using gradient inversion attacks (GIAs)~\cite{zhu2019deep, geiping2020inverting, dimitrov2022data}.

Since privacy is often a primary motivation for adopting FL, the threat posed by gradient inversion attacks should be taken seriously.
Recent research has highlighted the potential severity of such attacks, demonstrating successful reconstructions of high-resolution images~\cite{geiping2020inverting, geng2023improved}, feasibility at large batch sizes~\cite{geiping2020inverting, yin2021see}, and recovery of data from model updates accumulated over multiple local epochs~\cite{dimitrov2022data}.
However, many of these results rely on experimental assumptions such as architectural simplifications or idealized training conditions which differ substantially from those of modern, performance-optimized federated learning deployments.

This gap between attack demonstrations under controlled conditions and risk in realistic deployments has also been observed in other areas of federated learning security. For example, Shejwalkar et al.~\cite{shejwalkar2022back} present a systematic re-evaluation of poisoning attacks in federated learning and show that several attacks reported as highly effective in earlier work become substantially less impactful when evaluated under realistic training configurations and deployment constraints. For GIAs, numerous works have presented similar conclusions~\cite{guo2025exploring, du2025sok}. Their findings highlight the importance of distinguishing between diagnostic attack settings designed to expose worst-case vulnerabilities and conditions representative of production systems. Our work follows a similar philosophy in the context of gradient inversion, focusing on practical feasibility rather than upper-bound attack demonstrations.

Beyond academic exploration, the privacy risks posed by GIAs are increasingly shaping real-world ML deployments, particularly in sensitive domains such as healthcare and finance. 
FL is often seen as a key enabler of data-driven innovation in these areas, enabling collaborative model training without centralized access to personal data.
However, when the privacy guarantees of FL are poorly understood, uncertainty can hinder or even derail high-impact initiatives.

A concrete example of this risk emerged in a Swedish pilot project where two healthcare providers used FL to improve prediction models for patient readmission after heart failure. 
Following a review, the Swedish Authority for Privacy Protection (IMY) concluded that, under the project’s conditions, local model weights may contain sensitive information that could potentially be exposed via GIAs. 
In its public report~\cite{imy2023_pilotprojekt_dataskydd}, IMY emphasized that while such attacks are non-trivial, they could nevertheless enable reconstruction of private information from another participant in the federation. 
As a result, the project was not extended, despite its clear potential to improve healthcare delivery. More recently, however, the Swedish regulatory posture has shifted: privacy-enhancing technologies, including federated learning, are increasingly framed as enablers of responsible data sharing rather than as inherent liabilities, provided the specific privacy risks are adequately assessed~\cite{sou2026_44_integritetsframjande}. This evolving stance underscores the central problem we address: decisions to adopt or abandon FL hinge on risk assessments that are only as good as our understanding of when gradient inversion is actually feasible, and premature conclusions in either direction, over-fear or false reassurance, carry real costs. Although that project concerned tabular prediction models, it illustrates a general dependency of adoption decisions on risk assessment; we study vision models, where gradient inversion has been demonstrated most convincingly, and whether our conclusions transfer to other modalities remains open.

This example underscores a broader issue: in the absence of rigorous and transparent privacy risk assessments, promising AI systems risk being prematurely abandoned.
At the same time, accurately assessing such risks requires a clear understanding of the conditions under which privacy attacks are actually feasible in modern machine learning (ML) systems.

Motivated by these concerns, we present a comprehensive and realism-driven evaluation of the privacy risks posed by optimization-based GIAs in federated learning, using novel techniques to gain new insights into the robustness of different architectures and training setups while approaching settings expected in real world deployments. Our goal is to develop tools that assess GIA privacy risk beyond presenting failed or successful attack setups, increasing confidence in our privacy assessments.

\noindent\textbf{Contributions.} This paper makes four contributions, positioned against prior work in Section~\ref{subsec:sota}:
\begin{enumerate}
    \item \textbf{A target-interpolation feasibility probe:} a lightweight diagnostic that tests whether the attack objective singles out the direction toward the target, grading vulnerability at a fraction of the cost of a full attack.
    \item \textbf{A stronger training-mode attack:} exact recovery of the client's normalization statistics from shared running statistics, under the same assumptions as prior methods.
    \item \textbf{The first evaluation of gradient inversion on object detection:} YOLOv8-nano, RetinaNet, and DETR all resist under training-mode updates.
    \item \textbf{A mechanistic evaluation across modern architectures:} ResNet, Swin, SwinV2, ConvNeXt, MaxViT, and ViT on CIFAR-10 and ImageNet, tracing feasibility to normalization, stem design, resolution alignment, and update mode.
\end{enumerate}

\section{Systematization of GIAs}
\label{sec:sok}

In this section we recall the gradient inversion objective, situate the main families of attacks proposed in the literature, and motivate us in this study.

\subsection{The GIA objective}
\label{subsec:giabk}

Gradient inversion was first introduced by Zhu et al.~\cite{zhu2019deep}, who showed that private training data could, under certain conditions, be reconstructed from shared gradients. Their public implementation targeted relatively simple architectures such as LeNet, using sigmoid activations and shallow convolutional depth. The attack is formulated as an optimization problem,
\begin{equation}\label{eq:adv_obj}
\arg\min_{x, y} \; d\!\big( \nabla_\theta \mathcal{L}_\theta(x, y), \; \nabla_\theta \mathcal{L}_\theta(x^*, y^*) \big),
\end{equation}
where $(x^*, y^*)$ is the true input--label pair used in the client's gradient update $\nabla_\theta \mathcal{L}_\theta(x^*, y^*)$, $(x, y)$ is the attacker's reconstruction, $\theta$ are the model weights, $\mathcal{L}$ is the training loss, and $d(\cdot, \cdot)$ is a distance metric such as the Euclidean norm or cosine similarity.

Concretely, the attacker begins from an initial guess of the client's data which can be either randomly initialized~\cite{zhu2019deep, geiping2020inverting} or informed by prior knowledge of the data distribution~\cite{hatamizadeh2023gradient, yue2023gradient} and uses the global model to produce a synthetic update. It then iteratively adjusts the guessed input by gradient descent to minimize the distance between the synthetic and observed updates, nudging the reconstruction toward one that reproduces the client's gradient. Building on this formulation, Geiping et al.~\cite{geiping2020inverting} substantially improved reconstruction quality through total-variation priors, and demonstrated attacks on higher-resolution images on ResNet architectures. Notably, this core objective has remained essentially unchanged since its introduction: recent systematization efforts~\cite{carletti2025sok} observe that modern attacks continue to rely on the same gradient-matching formulation, differing primarily in threat model and auxiliary assumptions rather than in the underlying inversion mechanism.

\subsection{Attack families and scope}
\label{subsec:families}

Prior work is commonly organized into three families. \emph{Optimization-based} attacks (OP-GIA) follow Equation~\ref{eq:adv_obj} directly, recovering inputs by matching gradients through iterative optimization~\cite{zhu2019deep, geiping2020inverting}; subsequent refinements add latent-space priors~\cite{yue2023gradient}, GAN-based regularization~\cite{fang2023gifd}, or permutation-invariant objectives for aggregating updates across local steps~\cite{dimitrov2022data}. \emph{Generation-based} attacks (GEN-GIA) constrain the search to the manifold of a generative model trained on the data distribution, which can improve fidelity but presupposes a generator exposed to data representative of the client's~\cite{fang2023gifd}. \emph{Analytics-based} attacks (ANA-GIA) recover inputs in closed form by exploiting specific structures, such as an injected wide linear layer, and typically require the server to modify the model or training procedure~\cite{kariyappa2023cocktail}.

We focus on OP-GIA, hereafter referred to as just GIA, for reasons of both threat model and practicality. Generation- and latent-based attacks derive much of their strength from an auxiliary prior trained on the client data distribution; rather than granting the attacker this side information, we bound the contribution of such priors with an idealized prior in Section~\ref{sec:measure}. Analytics-based attacks, along with the broader class of attacks that assume an actively malicious server, for example model tampering in parameter-efficient fine-tuning~\cite{sami2025gradient}, the Cocktail Party attack~\cite{kariyappa2023cocktail}, and stealthy model manipulation all fall outside the honest-but-curious model we adopt (Section~\ref{sec:threat}), and such manipulations have moreover been shown to be detectable at the client~\cite{carletti2026detectability}. This leaves OP-GIA as the attack most consistent with our threat model. Attacks that add learned priors can be considerably stronger: GradViT~\cite{hatamizadeh2022gradvit} reconstructs $224\times224$ images from vision-transformer gradients by regularizing toward the batch-normalization statistics of a pretrained CNN, and reports vision transformers to be more vulnerable than CNNs. Such a prior encodes distribution-level knowledge of the client data, which a server in a privacy-sensitive federation may not possess, and our conclusions, which exclude learned priors, do not extend to it.

\subsection{Training dynamics, stochasticity, and normalization effects}
Several works study how training dynamics and stochastic components affect gradient inversion. Yin et al.~\cite{yin2020dreaming} introduced a batch-norm feature regularizer, later adapted in Yin et al.~\cite{yin2021see} to reconstruct batches from clients training with active BatchNorm layers. Huang et al.~\cite{huang2021evaluating} provide the most complete reproducible study of training-mode BatchNorm attacks and defenses. Hatamizadeh et al.~\cite{hatamizadeh2023gradient} report training-mode attacks in medical imaging, though their reconstructions rely heavily on domain-specific structural priors that do not readily transfer to natural images. More recently, Scheliga et al.~\cite{scheliga2023dropout} show that dropout does not constitute a reliable defense against gradient inversion, since dropout masks can be inferred from small-batch updates, a finding we adopt by disabling dropout in our own evaluation. Their experiments use a ViT-B/16 configuration that differs from the canonical ImageNet setup we study, so their results and ours are complementary rather than conflicting.

\subsection{Relation to recent systematizations and re-evaluations}
\label{subsec:relation}

On the architectural axis, the closest work is Zhang et al.~\cite{zhang2024does}, who study how model design shapes vulnerability across membership inference, model inversion, and gradient inversion for CNNs and transformers. Their analysis is an important step toward understanding privacy risk at the level of architecture rather than individual attacks; our work complements it by evaluating models in their intended data regimes and by extending the analysis to object detection, a setting their study does not consider.

Close in motivation are recent systematizations and re-evaluations of gradient inversion. Du et al.~\cite{du2025sok} and Carletti et al.~\cite{carletti2025sok} systematize the field along practical-effectiveness axes and threat-model taxonomies respectively, and Guo et al.~\cite{guo2025exploring} evaluate the attack families experimentally. We share their central conclusion, that gradient inversion is substantially more constrained in realistic settings than early demonstrations imply, and build on it by examining when and why attacks fail across architectures, training setups, and tasks.

\subsection{Gradient inversion beyond image classification}
\label{subsec:beyondcls}

Within vision, the gradient inversion literature has focused almost exclusively on image classification: from the original attack~\cite{zhu2019deep} through recent systematizations and re-evaluations~\cite{du2025sok, carletti2025sok, guo2025exploring}, the standard targets are classifiers such as LeNet, ResNet, and ViT on CIFAR-10 and ImageNet. But dense prediction tasks such as object detection remain, to our knowledge, unstudied. This gap matters because detection is widely deployed in the privacy-sensitive domains where federated learning is most attractive, and because detectors differ from classifiers in ways that plausibly affect leakage: structured multi-object outputs, composite losses over classification and box regression, and multi-scale feature pyramids that mix information across resolutions. We present, to our knowledge, the first evaluation of gradient inversion on object detection pipelines (Section~\ref{sec:objectdetect}).

\subsection{GIA feasibility probing}
\label{subsec:probing}

Evaluating an objective along a linear path between two points is an established tool for studying loss landscapes, originally applied in weight space to characterize training difficulty~\cite{goodfellow2014qualitatively}. To our knowledge, it has not previously been applied in input space to assess gradient inversion feasibility. Most closely related is Du et al.'s Input-Gradient Smoothness Analysis (IGSA)~\cite{du2025sok}, which estimates vulnerability from the local smoothness of the input-gradient map at the true input. Our target-interpolation feasibility probe (Section~\ref{subsec:probe}) instead evaluates the gradient-matching objective along the full path from an uninformed initialization to the target, capturing the descent an attacker's optimizer must actually traverse.

Evaluating the objective along a single line has a known weakness: the path ends at the target by construction, so the objective can decrease along it even when the target direction is no more attractive than an arbitrary one. We therefore complement the line evaluation with a step-wise comparison against random directions, in the spirit of the random-direction baselines used to visualize loss landscapes~\cite{li2018visualizing}.

\subsection{Enhancing the State of the Art}
\label{subsec:sota}

Relative to the work above, this paper advances the state of the art in five ways:
\begin{itemize}
    \item \textbf{From attack outcomes to feasibility diagnostics.} Prior evaluations report whether an attack succeeds, and local measures such as IGSA~\cite{du2025sok} characterize the objective only at the true input. Our probe tests, along the full path from an uninformed start, whether the objective singles out the target direction against random competitors, using about $4{,}000$ forward--backward passes instead of $\sim\!10^4$ iterations per attack run and $\sim\!10^5$ with hyperparameter search. It grades resistance that a full attack reports only as failure, for instance separating RetinaNet from DETR, which both resist.
    \item \textbf{A stronger training-mode attack under unchanged assumptions.} Prior training-mode attacks regularize toward the client's shared running statistics~\cite{yin2021see, huang2021evaluating}, which at default momentum are a 90/10 blend dominated by the server's own previous broadcast. We invert the momentum update to recover the exact batch statistics the client normalized with, requiring no additional information, and show that a proxy construction exceeds the prior reference even when running statistics are withheld.
    \item \textbf{Gradient inversion beyond classification.} We provide, to our knowledge, the first evaluation on object detection, covering one-stage convolutional (YOLOv8-nano, RetinaNet) and transformer (DETR) detectors. We also identify a numerical pitfall: for YOLOv8-nano the objective varies by only $\sim\!10^{-10}$ along the path, so attacks in single precision fail spuriously and double precision is required, a previously unreported source of false negatives in feasibility evaluations.
    \item \textbf{Separating prior from gradient information.} Prior-based attacks~\cite{yue2023gradient, fang2023gifd} report better reconstructions but cannot separate information supplied by the prior from information extracted from the gradients. Our controlled initialization supplies a prior of known, tunable informativeness, giving an upper-bound analysis of what a prior can contribute.
    \item \textbf{Mechanisms rather than verdicts.} Complementing architecture-level studies~\cite{zhang2024does}, we evaluate canonical configurations in their intended data regimes and trace feasibility to normalization type and placement, stem design, resolution alignment, and update mode, distinguishing two failure mechanisms: an objective with no target-specific signal, and one whose signal is local but underdetermined.
\end{itemize}

\section{ Attack and Probe Methodology }

This section describes our threat model, the attack constructions used throughout the paper, and the target-interpolation feasibility probe.

\subsection{Threat Model}
\label{sec:threat}
We consider a standard honest-but-curious federated learning setting. A central server orchestrates training rounds and aggregates client updates, observing the gradients or weight deltas each client shares, but it follows the prescribed protocol: it does not modify the model, loss function, or training procedure, nor alter client-side execution. Within this setting the server may attempt to reconstruct private training data from the updates it legitimately receives. This reflects the primary privacy concern in real-world deployments, that shared updates may inadvertently leak sensitive information even when all participants behave correctly, and it is the assumption underlying most deployed federated systems. Attacks that require an actively malicious server, which modifies the model or training procedure to induce leakage, fall outside this model and are not our focus (see Section~\ref{subsec:families}). Beyond falling outside the honest-but-curious setting, such manipulations have been shown to introduce anomalies detectable by clients through model inspection or in the gradient space~\cite{du2025sok, carletti2025sok}, making them a separate and separately-mitigable threat class.

\subsection{Attack Method}
\label{subsec:attackmethod}
In this section we describe the gradient inversion attack methodology used throughout this work, and formalize the strongest realistic training-mode attack constructions considered in our evaluation.

All attack variants build on the standard gradient inversion objective in Equation~\ref{eq:adv_obj}, which has remained essentially unchanged since its introduction by Zhu et al.~\cite{zhu2019deep}. This objective still forms the foundation of state-of-the-art methods: despite a wide range of proposed alternatives, recent systematization efforts~\cite{carletti2025sok} show that modern attacks continue to rely on the same core gradient-matching formulation, differing primarily in adversary threat model and auxiliary assumptions rather than in the underlying inversion mechanism. Throughout, we assume full label knowledge (Section~\ref{subsec:enablers}), fixing labels to $y^*$ and optimizing only over the input. For brevity we write $g(x)=\nabla_\theta\mathcal{L}_\theta(x,y^*)$ for the synthetic gradient and $g^*=\nabla_\theta\mathcal{L}_\theta(x^*,y^*)$ for the observed client gradient.

Following Geiping et al.~\cite{geiping2020inverting}, we instantiate the distance $d(\cdot,\cdot)$ as cosine distance and add a total-variation prior $\mathcal{R}_{\mathrm{TV}}$ to encourage natural image structure, giving the base objective used for architectures without exploitable normalization statistics:
\begin{equation}\label{eq:attack_base}
\begin{split}
\hat{x} \;=\; \arg\min_{x}\;
& \left[\, 1 - \frac{\langle g(x),\, g^*\rangle}{\lVert g(x)\rVert_2\,\lVert g^*\rVert_2} \,\right] \\
& +\; \lambda_{\mathrm{TV}}\,\mathcal{R}_{\mathrm{TV}}(x).
\end{split}
\end{equation}
Optimization is performed with signed gradient descent on the reconstruction, as in~\cite{geiping2020inverting}; we optionally reweight the cosine term across layers, treated as a tuned hyperparameter (Appendix~\ref{app:attack_robustness}).

For architectures whose normalization exposes activation statistics, the attacker can additionally regularize the reconstruction so that its internal statistics match a set of reference statistics. For each normalization layer $l$, let $\mu_l(x)$ and $\sigma_l^2(x)$ denote the (biased) mean and variance of the activations induced by $x$, and let
$\big(\mu_l^{\mathrm{ref}},(\sigma_l^{\mathrm{ref}})^2\big)$ denote the reference statistics. We define the feature-matching term
\begin{equation}\label{eq:featreg}
\begin{split}
\mathcal{R}_{\mathrm{feat}}\!\left(x;\,\mu^{\mathrm{ref}},\sigma^{\mathrm{ref}}\right)
= \sum_{l} \Big( &\big\lVert \mu_l(x)-\mu_l^{\mathrm{ref}}\big\rVert_2 \\
&+ \big\lVert \sigma_l^2(x)-(\sigma_l^{\mathrm{ref}})^2\big\rVert_2 \Big),
\end{split}
\end{equation}
where for BatchNorm layers the statistics are per-channel vectors, and for per-sample schemes (LayerNorm and its variants) they are aggregated to a single per-layer value. This term is the standard feature-matching regularizer introduced for data-free image synthesis~\cite{yin2020dreaming} and subsequently adopted throughout the gradient inversion literature. The three constructions below share Equation~\ref{eq:featreg} and the objective around it, and differ only in how the reference statistics are obtained; all reduce to Equation~\ref{eq:attack_base} when $\lambda_{\mathrm{BN}}=0$.

To ensure that observed failures are not artifacts of attack configuration, initialization, or insufficient optimization effort, all attacks are evaluated using multiple re-runs, extensive hyperparameter exploration, and targeted attempts to identify vulnerable samples, as detailed in Appendix~\ref{app:attack_robustness}.

\subsubsection{Shared Running Statistics}
\label{subsec:runningstats}
Training-mode attacks take the reference statistics directly from the client's shared BatchNorm buffers~\cite{yin2021see,huang2021evaluating,li2022e2egi,ye2024high,guo2025exploring}. After a client update, the running statistics of layer $l$ are $\big(\mu_l^{(t)},(\sigma_l^{(t)})^2\big)$, and these are used as-is:
\begin{equation}\label{eq:attack_running}
\begin{split}
\hat{x} \;=\; \arg\min_{x}\;
& \left[\, 1 - \frac{\langle g(x),\, g^*\rangle}{\lVert g(x)\rVert_2\,\lVert g^*\rVert_2} \,\right] \\
& +\; \lambda_{\mathrm{TV}}\,\mathcal{R}_{\mathrm{TV}}(x) \\
& +\; \lambda_{\mathrm{BN}}\,\mathcal{R}_{\mathrm{feat}}\!\left(x;\,\mu^{(t)},\sigma^{(t)}\right).
\end{split}
\end{equation}
This is the construction we adopt as the reference point for prior work throughout our evaluation.

\subsubsection{Exact Batch Statistic Inference via Shared Running Statistics}
\label{subsec:sharedstats}
The running statistics of Equation~\ref{eq:attack_running} are not the statistics the client actually normalized with. Because they are updated with a momentum rule,
$\mu_l^{(t)} = (1-\rho)\,\mu_l^{(t-1)} + \rho\,\mu_l^{\mathrm{batch}}$, they are a blend of the client's batch statistics and the statistics the server itself broadcast in the previous round. Since the attacker observes both endpoints of this update, the blend can be inverted to recover the \emph{exact} per-channel batch statistics used in the client's forward and backward pass:
\begin{equation}\label{eq:momentum}
\begin{aligned}
\mu_l^{\mathrm{exact}} &= \frac{\mu_l^{(t)}-(1-\rho)\,\mu_l^{(t-1)}}{\rho}, \\[4pt]
(\sigma_l^{\mathrm{exact}})^2 &= \frac{n_l-1}{n_l}\cdot
\frac{(\sigma_l^{(t)})^2-(1-\rho)\,(\sigma_l^{(t-1)})^2}{\rho}.
\end{aligned}
\end{equation}
With the default momentum $\rho=0.1$, this reduces to
$\mu_l^{\mathrm{exact}} = 10\,\mu_l^{(t)} - 9\,\mu_l^{(t-1)}$. The factor
$(n_l-1)/n_l$, where $n_l$ is the number of elements per channel, converts the unbiased running variance back to the biased variance used during forward normalization, so that it matches the biased variance computed on the reconstruction in Equation~\ref{eq:featreg}. The inversion is exact for a single local step at known momentum, the setting we study throughout. Using these exact statistics as the reference gives the strongest training-mode attack applicable to architectures that expose running statistics:
\begin{equation}\label{eq:attack_exact}
\begin{split}
\hat{x} \;=\; \arg\min_{x}\;
& \left[\, 1 - \frac{\langle g(x),\, g^*\rangle}{\lVert g(x)\rVert_2\,\lVert g^*\rVert_2} \,\right] \\
& +\; \lambda_{\mathrm{TV}}\,\mathcal{R}_{\mathrm{TV}}(x) \\
& +\; \lambda_{\mathrm{BN}}\,\mathcal{R}_{\mathrm{feat}}\!\left(x;\,\mu^{\mathrm{exact}},\sigma^{\mathrm{exact}}\right).
\end{split}
\end{equation}
Anchoring the reconstruction to the true normalization behavior substantially reduces ambiguity in the inversion problem. This requires no information beyond what Equation~\ref{eq:attack_running} already assumes, since the earlier endpoint $\mu_l^{(t-1)}$ is the server's own broadcast: the two constructions differ in what is extracted from the observation, not in what is observed.

\subsubsection{Proxy Normalization Statistics}
\label{subsec:proxystats}
If running statistics are not shared or if the architecture uses per-sample normalization such as LayerNorm, the attacker has no direct access to the client's normalization behavior. In this more general and widely applicable setting, we estimate reference statistics from auxiliary data. Given a candidate proxy batch, we perform a single forward pass and capture the per-layer activation statistics $\mu_l^{\mathrm{proxy}},(\sigma_l^{\mathrm{proxy}})^2$ (per-channel for BatchNorm; aggregated per layer for LayerNorm-type layers), which then serve as the reference in Equation~\ref{eq:featreg}:
\begin{equation}\label{eq:attack_proxy}
\begin{split}
\hat{x} \;=\; \arg\min_{x}\;
& \left[\, 1 - \frac{\langle g(x),\, g^*\rangle}{\lVert g(x)\rVert_2\,\lVert g^*\rVert_2} \,\right] \\
& +\; \lambda_{\mathrm{TV}}\,\mathcal{R}_{\mathrm{TV}}(x) \\
& +\; \lambda_{\mathrm{BN}}\,\mathcal{R}_{\mathrm{feat}}\!\left(x;\,\mu^{\mathrm{proxy}},\sigma^{\mathrm{proxy}}\right).
\end{split}
\end{equation}
Since no single proxy batch is guaranteed to reflect the client's statistics, we draw several candidate proxy batches and select the one yielding the strongest reconstruction as part of the hyperparameter search (Appendix~\ref{app:attack_robustness}). This reflects a strong but realistic adversary who probes multiple plausible normalization estimates, and requires no privileged access to client-side information; it applies uniformly across BatchNorm and LayerNorm architectures.

\subsubsection{Auxiliary Attack Enablers and Scope}
\label{subsec:enablers}
Prior work shows that the two commonly cited practical obstacles, unknown labels and the presence of dropout, are not fundamental blockers for gradient inversion in the low-batch FedSGD regimes most relevant to our study. Label reconstruction is typically straightforward as long as client updates do not mix multiple examples from the same class~\cite{huang2021evaluating}. Likewise, dropout does not prevent inversion at very small batch sizes: recent results demonstrate that dropped activations (or equivalent masks) can be inferred reliably for batches below a few samples ($B<4$), enabling attacks to proceed with appropriate techniques~\cite{scheliga2023dropout}.

Our study focuses on the regime where performance-optimized modern vision architectures and ImageNet-scale inputs already cease to be vulnerable to GIAs, before dropout or label inference become limiting factors. Accordingly, we disable dropout and assume full label knowledge.

\subsection{The Target-Interpolation Feasibility Probe}
\label{subsec:probe}
Running a full gradient inversion attack is expensive, and a failed reconstruction
is ambiguous: it may indicate that the gradients genuinely lack recoverable
information, or merely that the optimizer failed to navigate a difficult landscape.
To disentangle these, and to estimate feasibility at a small fraction of the cost
of a full attack, we introduce a \emph{target-interpolation feasibility probe}.
Rather than optimizing a reconstruction, the probe walks a fixed path from an
uninformed starting point to the true input and tests, at each step, whether the
attack objective singles out the direction toward the target.

Let $x_0 \sim \mathcal{N}(0, I)$ be a random reference image and $x^*$ the client's true input, and define the interpolation path
\begin{equation}\label{eq:probe_path}
x(\alpha) \;=\; (1-\alpha)\,x_0 \;+\; \alpha\, x^*,
\end{equation}
so that $\alpha=0$ is pure noise and $\alpha=1$ is exactly the target. Throughout, $\mathcal{J}_{g^*}$ is the attack objective of Section~\ref{subsec:attackmethod} with the same client gradient $g^*$ (all probe results reported here use the gradient-matching term alone, so regularizers such as the total-variation prior cannot reward noise removal independently of the target; Appendix~\ref{app:probe}), evaluated without any optimization. That the objective decreases along the path is not by itself informative: every point is a mixture of noise and target, so moving along it removes noise while adding target content, and the objective may reward the former without preferring the target at all. It may also decrease toward the target yet decrease \emph{faster} in other directions, pulling a descent method away from the path.

We therefore compare, at each step, the target direction against random alternatives. For path points $x_i = x(\alpha_i)$ with step $\delta_i = x_{i+1} - x_i$, let
\begin{equation}\label{eq:spec_target}
\Delta^{*}_i \;=\; \mathcal{J}_{g^*}(x_i) - \mathcal{J}_{g^*}(x_i + \delta_i)
\end{equation}
be the decrease of the objective along the target step (positive is downhill), and compare it with $K$ random steps of the same length,
\begin{equation}\label{eq:spec_random}
\Delta_{k,i} \;=\; \mathcal{J}_{g^*}(x_i) - \mathcal{J}_{g^*}\!\big(x_i + \lVert\delta_i\rVert_2\, u_k\big), \qquad k = 1,\dots,K,
\end{equation}
where $u_k$ is a unit-norm \emph{smooth} random direction (Gaussian noise sampled at one quarter of the input resolution and bilinearly upsampled), a stronger competitor than pixel-level noise, which in high dimension is nearly orthogonal to any descent direction. Each step is summarized by whether the best competitor beats the target step,
\begin{equation}\label{eq:spec_stats}
b_i \;=\; \mathbb{1}\!\left[\max_k \Delta_{k,i} > \Delta^{*}_i\right],
\end{equation}
which in particular sets $b_i = 1$ wherever the target step itself goes uphill. We report the fraction of noise draws with $b_i = 1$ at each $\alpha$, and summarize it by its mean $\nu$ over $\alpha \in [0, 0.9]$, the \emph{non-specificity rate}, excluding the immediate neighbourhood of the target where $\Delta^*_i$ vanishes. A target-specific landscape has $\nu \approx 0$: no random step of the same size competes with the step toward the target. A non-specific landscape has $\nu$ close to $1$: the objective offers an optimizer starting from noise no target-specific signal to follow.

With $K=10$ at every fifth point of a 141-point path, the complete probe over ten noise draws takes about $4{,}000$ forward--backward passes, about an order of magnitude below a single attack run (whose $\sim\!10^4$ iterations each require a double backward pass) and far below an attack with hyperparameter search. Because step-wise differences can be many orders of magnitude smaller than the objective itself (for YOLOv8-nano in inference mode the objective spans only $\sim\!10^{-10}$ along the entire path), the cosine term is evaluated in double precision.

The probe evaluates a one-dimensional slice of a high-dimensional landscape against a finite sample of competing directions, so it is a screen rather than a proof of (in)feasibility: $K$ random directions only lower-bound how well the best competitor does, and the optimizer is not confined to the path. We therefore treat target-specific descent as an empirically necessary condition. Across our experiments, no configuration the probe flags as non-specific is reconstructable (Sections~\ref{sec:objectdetect}, \ref{sec:measure}), while ViT-B/16 and ConvNeXt-T show that passing the test does not guarantee reconstruction.

\section{Attack Validation and Architectural Dependence}
\label{sec:validation}

We first validate our attack on ResNet18 variants with active BatchNorm, a standard benchmark for training-mode GIA~\cite{huang2021evaluating,yin2021see, guo2025exploring, ye2024high}, to establish that later negative results are not artifacts of a weak attack; the same experiment also reveals an initial architectural dependence of feasibility. We compare the two constructions of Section~\ref{subsec:attackmethod}, exact batch statistics recovered from shared running statistics (Eq.~\ref{eq:attack_exact}) and proxy statistics when running statistics are withheld (Eq.~\ref{eq:attack_proxy}), against the reference construction of prior methods (Eq.~\ref{eq:attack_running}).

We distinguish \emph{pre-activation} and \emph{post-activation} ResNets. In pre-activation ResNets~\cite{he2016identity}, each residual unit applies BatchNorm and ReLU before the convolution (BN → ReLU → conv), leaving the skip connection as a pure identity. In the original post-activation design~\cite{he2016resnet}, normalization and activation follow the convolution (conv → BN → ReLU), with an additional ReLU after the residual addition.

\begin{figure*}[t]
\centering
\setlength{\tabcolsep}{6pt}
\renewcommand{\arraystretch}{1.1}

\begin{tabular}{c c c c c}
 & \multicolumn{2}{c}{\textbf{Pre-activation ResNet18}}
 & \multicolumn{2}{c}{\textbf{Post-activation ResNet18}} \\
\cmidrule(lr){2-3} \cmidrule(lr){4-5}
 & \textbf{Original} & \textbf{Reconstruction}
 & \textbf{Original} & \textbf{Reconstruction} \\
\midrule

\textbf{Ours (shared)} &
\includegraphics[width=0.19\textwidth]{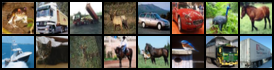} &
\includegraphics[width=0.19\textwidth]{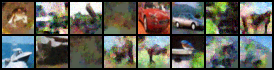} &
\includegraphics[width=0.19\textwidth]{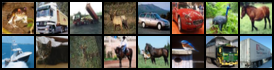} &
\includegraphics[width=0.19\textwidth]{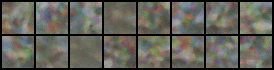} \\
& \multicolumn{4}{c}{%
\makebox[0.76\textwidth]{SSIM=0.054 \hspace{14em} SSIM=0.041}
} \\[6pt]

\textbf{Ours (no shared)} &
\includegraphics[width=0.19\textwidth]{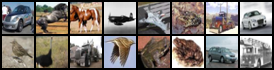} &
\includegraphics[width=0.19\textwidth]{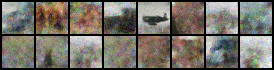} &
\includegraphics[width=0.19\textwidth]{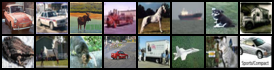} &
\includegraphics[width=0.19\textwidth]{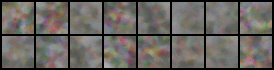} \\

& \multicolumn{4}{c}{%
\makebox[0.76\textwidth]{SSIM=0.040 \hspace{14em} SSIM=0.024}
} \\[6pt]

\textbf{Prior methods} &
\includegraphics[width=0.19\textwidth]{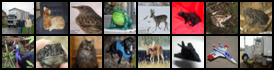} &
\includegraphics[width=0.19\textwidth]{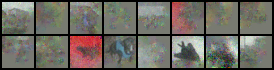} &
\includegraphics[width=0.19\textwidth]{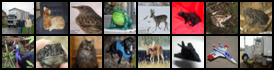} &
\includegraphics[width=0.19\textwidth]{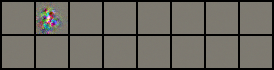} \\
& \multicolumn{4}{c}{%
\makebox[0.76\textwidth]{SSIM=0.030 \hspace{14em} SSIM=0.013}
} \\[6pt]

\end{tabular}

\caption{
\textbf{Validation of gradient inversion under training-mode BatchNorm (SSIM $\uparrow$).}
Columns compare pre-activation and post-activation ResNet18 architectures, while rows vary the reference statistics used to guide the reconstruction: exact batch statistics recovered from shared running statistics (Equation~\ref{eq:attack_exact}), proxy statistics estimated from auxiliary data when running statistics are withheld (Equation~\ref{eq:attack_proxy}), and the running-statistic reference used by prior training-mode methods (Equation~\ref{eq:attack_running}).
The post-activation version of the model consistently resists the GIA which highlights a strong architectural dependence of attack feasibility.
}
\label{fig:method_validation_bn_training_mode}
\end{figure*}

Figure~\ref{fig:method_validation_bn_training_mode} shows that this seemingly minor choice decides feasibility. Gradient inversion succeeds consistently on pre-activation ResNet18, yielding reconstructions that retain semantic structure, but fails on post-activation ResNet18, whose reconstructions lack meaningful semantic content, although the two models are identical in depth, width, and capacity. Since SSIM separates these regimes only weakly at this resolution, success is judged jointly with visual inspection (Appendix~\ref{app:attack_robustness}). The difference is tied to training-mode normalization: with clients in inference mode, where BatchNorm becomes affine, post-activation ResNet18 becomes reconstructable (Table~\ref{tab:gia_summary}), the same mechanism we observe for object detectors in Section~\ref{sec:objectdetect}.

Comparing the three reference-statistic constructions, inverting the momentum update on pre-activation ResNet18 improves reconstruction quality over using the running statistics directly (SSIM $0.054$ versus $0.030$). With running statistics withheld, the proxy construction still exceeds the prior-method reference ($0.040$ versus $0.030$), although its proxy batch is selected for reconstruction quality and it is therefore best read as a strong-adversary estimate. The same ordering holds on post-activation ResNet18 ($0.041$, $0.024$, $0.013$), although all three reconstructions are semantically empty there.

\noindent\textbf{Key findings.}
\begin{itemize}
    \item Our attack is at least as strong as the training-mode construction of prior methods, and remains stronger without access to the client's running statistics, so the negative results in later sections are not artifacts of a weak attack.
    \item Seemingly minor design choices decide feasibility: reordering the residual unit from pre- to post-activation is enough to make training-mode updates resist inversion, while inference-mode updates make the post-activation model vulnerable again.
\end{itemize}

\section{Gradient Inversion on Object Detection}
\label{sec:objectdetect}

We evaluate three architectures spanning the dominant detector families: YOLOv8-nano (one-stage, CSP-based convolutional), RetinaNet (one-stage, ResNet-FPN), and DETR (transformer), on COCO~\cite{lin2014coco}, under favorable attacker assumptions throughout: batch size one and full label knowledge. The detectors also differ in what their realistic training mode is. YOLOv8 trains with active BatchNorm, whereas RetinaNet and DETR are conventionally fine-tuned with frozen BatchNorm; for RetinaNet we additionally evaluate trainable BatchNorm, and for DETR, whose backbone normalization is frozen and whose transformer uses LayerNorm, training and inference mode coincide.

\subsection{Training-mode updates resist inversion across all detectors}
With client updates produced in training mode, gradient inversion fails for all three detectors, even at batch size one: reconstructions collapse to noise or weak global colour, with no recoverable object-level structure. We omit these images, as uniform noise across all configurations is uninformative. Since this holds across the convolutional, ResNet-FPN, and transformer families alike, resistance appears to be characteristic of detection pipelines rather than of a single architectural style.

\begin{figure}[t]
    \centering
    \begin{subfigure}[b]{0.48\columnwidth}
        \includegraphics[width=\linewidth]{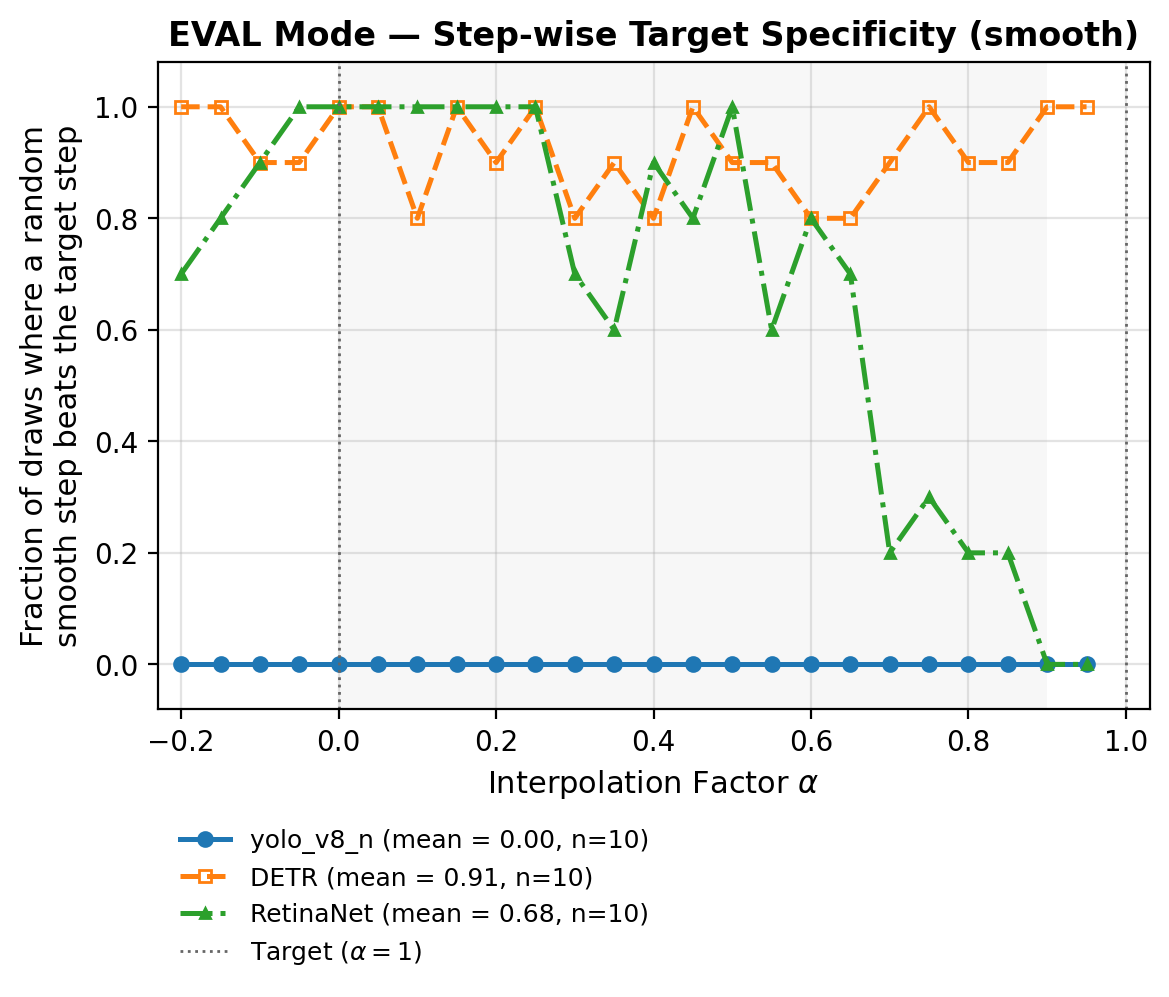}
        \caption{Inference mode}
        \label{fig:od_probe_eval}
    \end{subfigure}\hfill
    \begin{subfigure}[b]{0.48\columnwidth}
        \includegraphics[width=\linewidth]{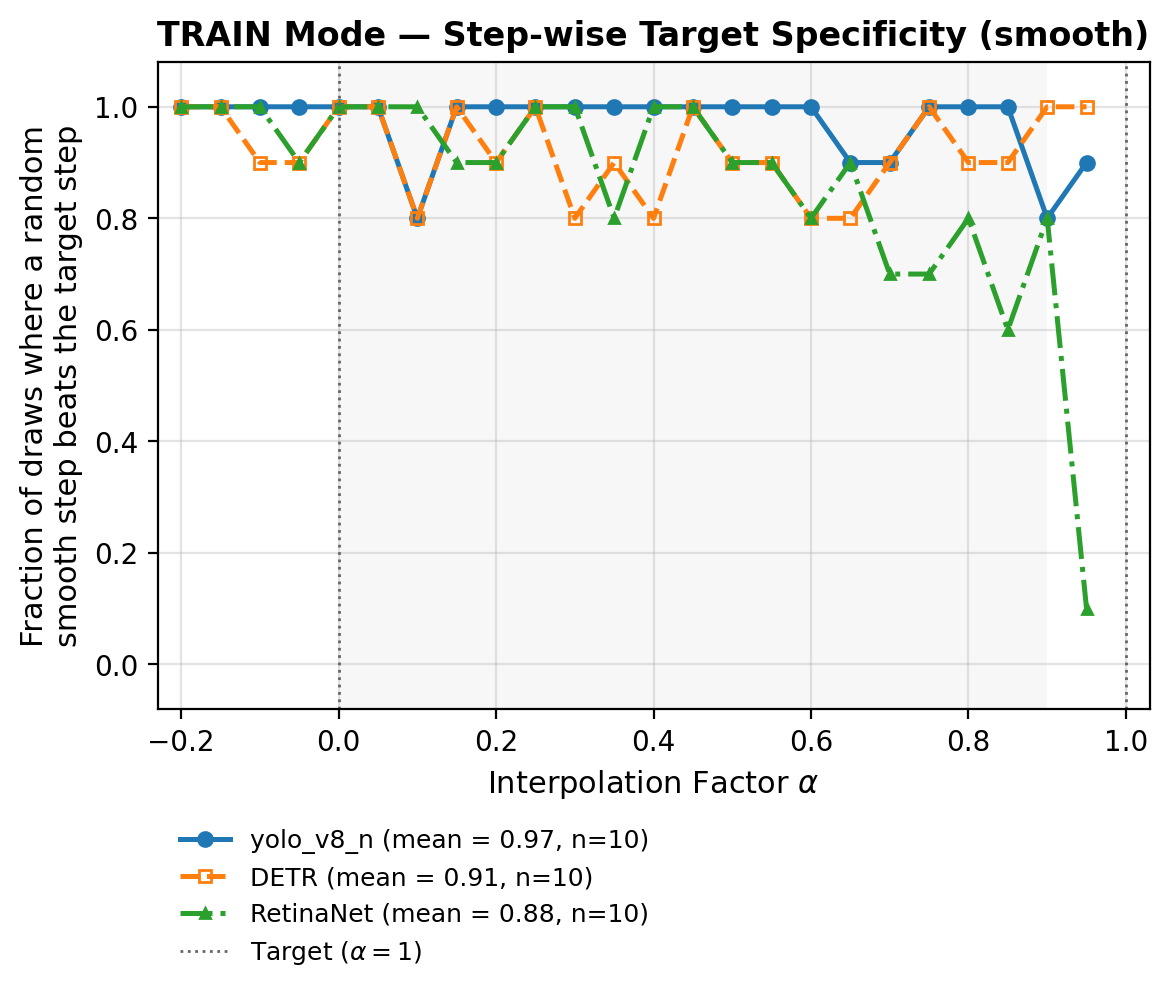}
        \caption{Training mode}
        \label{fig:od_probe_train}
    \end{subfigure}
    \caption{\textbf{Feasibility probe on object detectors} (gradient-matching objective, $n{=}10$ noise draws, $K{=}10$ smooth directions): fraction of draws in which a random step beats the step toward the target along the noise$\rightarrow$target path; the legend gives $\nu$ (lower is more target-specific). (a) In inference mode, the ordering YOLOv8-nano $<$ RetinaNet $<$ DETR matches the full-attack outcomes in Fig.~\ref{fig:od_recon}. (b) In training mode, all three detectors are non-specific.}
    \label{fig:od_probe}
\end{figure}

The feasibility probe (Fig.~\ref{fig:od_probe_train}) shows why. In training mode the objective offers no target-specific direction for any detector: the non-specificity rate is $\nu = 0.97$ for YOLOv8-nano, $0.88$ for RetinaNet, and $0.91$ for DETR, and the step toward the target itself goes uphill at $52\%$, $36\%$, and $47\%$ of path points, respectively. An optimizer starting from noise therefore has no consistent signal to follow, consistent with the failed attacks.

\subsection{Inference mode can enable reconstruction}

\begin{figure}[t]
    \centering
    \begin{subfigure}[b]{0.24\columnwidth}
        \includegraphics[width=\linewidth]{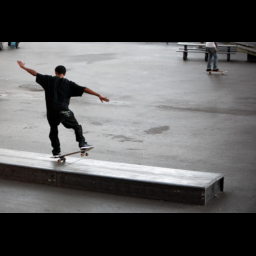}
        \caption{Original \\ . \\ .}
    \end{subfigure}\hfill
    \begin{subfigure}[b]{0.24\columnwidth}
        \includegraphics[width=\linewidth]{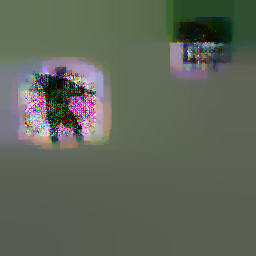}
        \caption{YOLOv8-n\\ SSIM=0.2978 / LPIPS=0.6414}
    \end{subfigure}\hfill
    \begin{subfigure}[b]{0.24\columnwidth}
        \includegraphics[width=\linewidth]{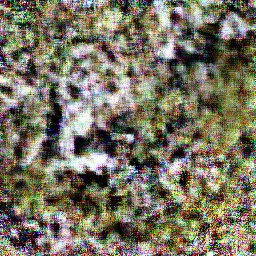}
        \caption{RetinaNet\\ SSIM=0.0256 / LPIPS=0.7600}
    \end{subfigure}\hfill
    \begin{subfigure}[b]{0.24\columnwidth}
        \includegraphics[width=\linewidth]{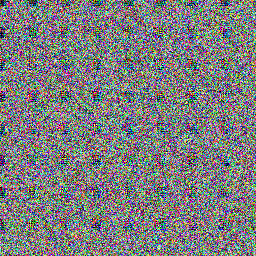}
        \caption{DETR\\ SSIM=0.0060 / LPIPS=1.4869}
    \end{subfigure}
    \caption{\textbf{Gradient inversion on object detectors with inference-mode client updates.} Only YOLOv8-nano yields faint but visible structure, and only under this unrealistic setting; RetinaNet and DETR collapse to noise. None of the three is reconstructable under training-mode updates (Fig.~\ref{fig:od_probe}).}
    \label{fig:od_recon}
\end{figure}

In inference mode, YOLOv8-nano becomes reconstructable, yielding faint but visible structure, while RetinaNet and DETR continue to resist (Fig.~\ref{fig:od_recon}). This result requires double precision: the objective varies by only $\sim\!10^{-10}$ along the path, and in single precision the attack collapses to a uniform image.

We attribute this to normalization. In training mode, a BatchNorm layer normalizes each activation with statistics computed from the current input, even at batch size one, making the backward pass a non-linear function of the input that couples the gradient to the very quantity the attacker is trying to recover. In inference mode these statistics are replaced by fixed running estimates, so normalization becomes affine and the backward pass is correspondingly better conditioned.

The probe grades this per architecture (Fig.~\ref{fig:od_probe_eval}). For YOLOv8-nano the objective is target-specific along the entire path ($\nu = 0.00$): across all ten noise draws, no random step ever lowers it as much as the step toward the target, consistent with its successful reconstruction. DETR is unchanged from training mode ($\nu = 0.91$), as expected since its two modes coincide. RetinaNet falls between the two: its objective is non-specific over most of the path ($\nu = 0.68$) but becomes target-specific in a basin around the target ($\alpha \gtrsim 0.7$). So although a full attack reports it, like DETR, as resistant, its landscape is markedly closer to feasibility, a distinction a binary attack outcome cannot make.

\subsection{Implications}
\begin{itemize}
    \item Object detectors resist gradient inversion under training-mode updates even at batch size one, the most favorable batch setting for an attacker, across one-stage convolutional (YOLOv8-nano, RetinaNet) and transformer (DETR) architectures.
    \item The one vulnerability we observe requires inference-mode updates on YOLOv8-nano, a configuration that does not match how it is trained in practice; RetinaNet and DETR, whose conventional training uses frozen normalization, resist in either mode.
\end{itemize}

\section{Measuring ``Attack Difficulty'' via Controlled Initialization}
\label{sec:measure}

The attacks above all start from an uninformed initialization. Attacks that strengthen gradient inversion with informative priors~\cite{yue2023gradient, fang2023gifd, hatamizadeh2023gradient} report better reconstructions, but cannot separate information supplied by the prior from information extracted from the gradients. We isolate the two by giving the attacker a prior of known, tunable informativeness: the target itself, corrupted by a controlled amount of noise.

\newcommand{\imgcell}[2]{\begin{minipage}[c]{0.075\textwidth}\centering\includegraphics[width=\linewidth]{#1}\\[-2pt]{\scriptsize LPIPS=#2}\end{minipage}}
\newcommand{\startcell}[2]{\begin{minipage}[c]{0.075\textwidth}\centering\includegraphics[width=\linewidth]{#1}\\[-2pt]{\scriptsize LPIPS=#2}\end{minipage}}
\begin{figure*}
[t]\centering
\adjustbox{max width=\textwidth}{%
\begin{tabular}{lccccccccccc}
\toprule
\textbf{Starting noise factor} & 0.0 & 0.1 & 0.2 & 0.3 & 0.4 & 0.5 & 0.6 & 0.7 & 0.8 & 0.9 & 1.0 \\ \midrule

\textbf{Starting point} &
\startcell{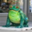}{0.000} &
\startcell{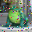}{0.028} &
\startcell{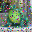}{0.064} &
\startcell{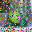}{0.135} &
\startcell{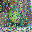}{0.200} &
\startcell{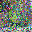}{0.267} &
\startcell{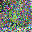}{0.358} &
\startcell{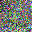}{0.463} &
\startcell{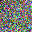}{0.498} &
\startcell{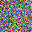}{0.513} &
\startcell{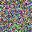}{0.441} \\
\midrule
\textbf{MaxViT-T} &
\imgcell{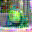}{0.014} &
\imgcell{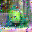}{0.018} &
\imgcell{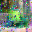}{0.022} &
\imgcell{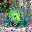}{0.056} &
\imgcell{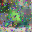}{0.076} &
\imgcell{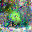}{0.070} &
\imgcell{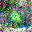}{0.111} &
\imgcell{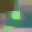}{0.087} &
\imgcell{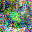}{0.105} &
\imgcell{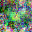}{0.106} &
\imgcell{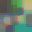}{0.131} \\
\midrule
\textbf{ViT-B/16} &
\imgcell{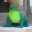}{0.017} &
\imgcell{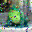}{0.015} &
\imgcell{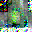}{0.061} &
\imgcell{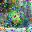}{0.070} &
\imgcell{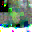}{0.048} &
\imgcell{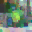}{0.037} &
\imgcell{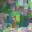}{0.079} &
\imgcell{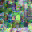}{0.097} &
\imgcell{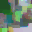}{0.066} &
\imgcell{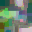}{0.077} &
\imgcell{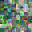}{0.170} \\
\midrule
\textbf{ConvNeXt-T} &
\imgcell{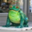}{0.000} &
\imgcell{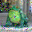}{0.009} &
\imgcell{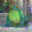}{0.015} &
\imgcell{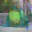}{0.020} &
\imgcell{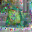}{0.047} &
\imgcell{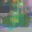}{0.069} &
\imgcell{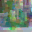}{0.063} &
\imgcell{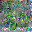}{0.186} &
\imgcell{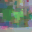}{0.076} &
\imgcell{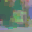}{0.071} &
\imgcell{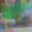}{0.091} \\
\midrule
\textbf{SwinV2-T} &
\imgcell{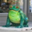}{0.000} &
\imgcell{results/cifar_noise_study/swinv2_1.0_0.9950.png}{0.000} &
\imgcell{results/cifar_noise_study/swinv2_1.0_0.9950.png}{0.000} &
\imgcell{results/cifar_noise_study/swinv2_1.0_0.9950.png}{0.000} &
\imgcell{results/cifar_noise_study/swinv2_1.0_0.9950.png}{0.000} &
\imgcell{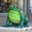}{0.000} &
\imgcell{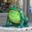}{0.000} &
\imgcell{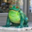}{0.000} &
\imgcell{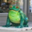}{0.000} &
\imgcell{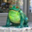}{0.001} &
\imgcell{results/cifar_noise_study/swinv2_1.0_0.9950.png}{0.000} \\
\midrule
\textbf{Swin-T} &
\imgcell{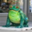}{0.000} &
\imgcell{results/cifar_noise_study/swin_0.5_1.0000.png}{0.000} &
\imgcell{results/cifar_noise_study/swin_0.5_1.0000.png}{0.000} &
\imgcell{results/cifar_noise_study/swin_0.5_1.0000.png}{0.000} &
\imgcell{results/cifar_noise_study/swin_0.5_1.0000.png}{0.000} &
\imgcell{results/cifar_noise_study/swin_0.5_1.0000.png}{0.000} &
\imgcell{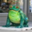}{0.000} &
\imgcell{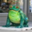}{0.000} &
\imgcell{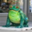}{0.000} &
\imgcell{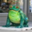}{0.000} &
\imgcell{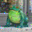}{0.003} \\
\\ \bottomrule
\end{tabular}%
}

\caption{\textbf{Controlled initialization on CIFAR-10 (LPIPS $\downarrow$).}
Each row initializes the attack from the target corrupted by increasing noise (factor $0.0$ = target, $1.0$ = pure noise; top row shows the starting points) and optimizes to convergence. Resistant architectures (ViT-B/16, MaxViT-T, ConvNeXt-T) recover semantic content only when the initialization is already close to the target, whereas Swin-T and SwinV2-T reconstruct accurately from any start.}

\label{fig:cifar-noise-grid}
\end{figure*}

Concretely, we run the full attack from points on the probe's interpolation path (Section~\ref{subsec:probe}), initializing at $x(\alpha) = (1-\alpha)\,x_0 + \alpha\,x^*$ and optimizing to convergence. A noise factor of $0.0$ corresponds to initialization at the true image and $1.0$ to pure noise. Knowing the target's pixels, even partially corrupted, is far stronger than any realistic prior, so this setting serves as a diagnostic upper bound rather than an attack.

\paragraph{Architectural scope.} The models are ImageNet architectures adapted for CIFAR-10: early-stage downsampling in each stem is reduced so that spatial information survives at 32×32, which is required for functional training (for ConvNeXt-T, replacing the 4×4 stride-4 stem with a 3×3 stride-1 convolution improves accuracy by roughly 13 percentage points). These adapted stems can themselves introduce vulnerabilities; we evaluate the canonical configurations in Section~\ref{sec:evalimagenet}.

Figure~\ref{fig:cifar-noise-grid} shows a strongly architecture-dependent effect. Swin-T and SwinV2-T, already vulnerable from an uninformed start, reconstruct accurately from any initialization. For ViT-B/16, MaxViT-T, and ConvNeXt-T, final quality instead degrades steadily as the initialization becomes less informative (e.g., MaxViT-T from LPIPS $0.014$ to $0.131$): the reconstruction is only as good as its starting point. These models still improve substantially over their noisy initializations, but we attribute much of this to the total-variation prior, which removes noise from any starting image regardless of whether the gradient-matching term carries target information.

\begin{figure}[t]
    \centering
    \includegraphics[width=0.7\linewidth]{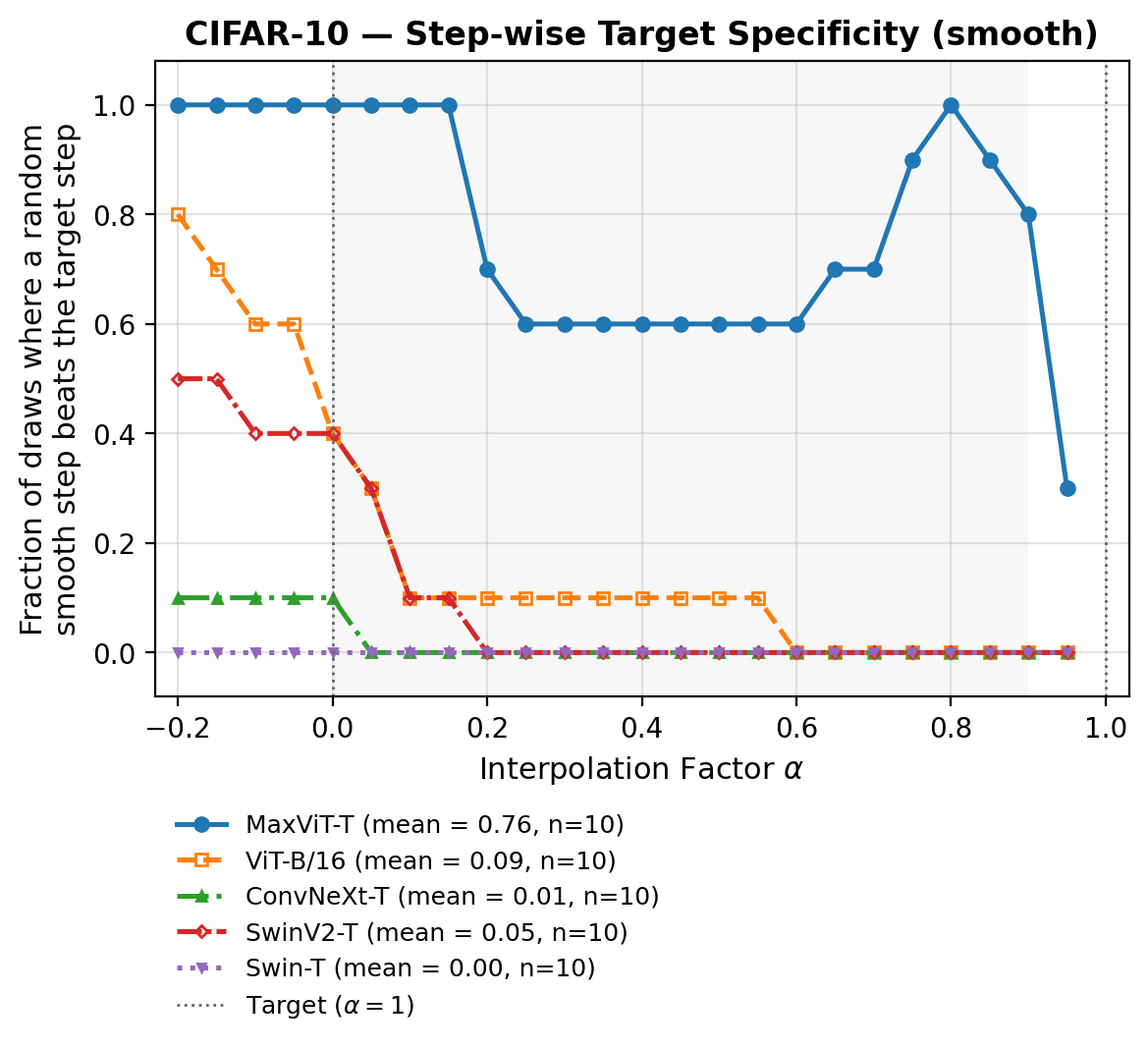}
    \caption{\textbf{Feasibility probe on the CIFAR-10 adapted models of Fig.~\ref{fig:cifar-noise-grid}} (gradient-matching objective, $n{=}10$ noise draws): fraction of draws in which the best of $K{=}10$ random smooth steps lowers the objective more than the step toward the target, with the mean over $\alpha\in[0,0.9]$ in the legend. Only MaxViT-T is non-specific along most of the path.}
    \label{fig:cifar_probe}
\end{figure}

The probe (Fig.~\ref{fig:cifar_probe}) separates the resistant models further. MaxViT-T is non-specific along most of the path ($\nu = 0.76$), and near the noise end the target step goes uphill. For ViT-B/16 ($\nu = 0.09$) and ConvNeXt-T ($\nu = 0.01$), by contrast, the step toward the target is almost always the steepest of the tested directions, as it is for SwinV2-T ($0.05$) and Swin-T ($0.00$). Their resistance is therefore not caused by competing directions along the path. This illustrates the probe's one-sided nature: target-specific descent is empirically necessary for feasibility but, as these two models show, not sufficient, since the optimizer is not confined to the path.

\noindent\textbf{Key findings.}
\begin{itemize}
    \item The benefit of an informed prior is bounded by the information the gradients already expose: where it is present (Swin-T, SwinV2-T), reconstruction succeeds from any start; where it is absent (ViT-B/16, MaxViT-T, ConvNeXt-T), reconstruction quality tracks the initialization rather than the gradients.
    \item This supports, though does not by itself establish, the broader reading that gradient inversion is limited more by the information content of the gradients than by attack optimization.
\end{itemize}

\section{Evaluation on Modern ImageNet Models}
\label{sec:evalimagenet}

We next evaluate modern vision architectures trained on ImageNet in their canonical configurations at $224\times224$ resolution, under identical and favorable attack conditions: batch size one, full label knowledge, dropout disabled, and the proxy-statistic construction of Eq.~\ref{eq:attack_proxy}.

\begin{figure*}[t]
    \centering

    \begin{subfigure}{0.16\textwidth}
        \centering
        \includegraphics[height=2.5cm]{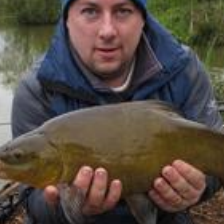}
        \caption{Original\\.}
    \end{subfigure}
    \hfill
    \begin{subfigure}{0.16\textwidth}
        \centering
        \includegraphics[height=2.5cm]{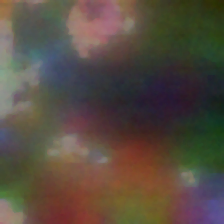}
        \caption{Swin-T\\(SSIM=0.387)}
    \end{subfigure}
    \hfill
    \begin{subfigure}{0.16\textwidth}
        \centering
        \includegraphics[height=2.5cm]{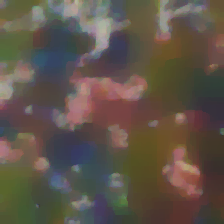}
        \caption{SwinV2-T\\(SSIM=0.338)}
    \end{subfigure}
    \hfill
    \begin{subfigure}{0.16\textwidth}
        \centering
        \includegraphics[height=2.5cm]{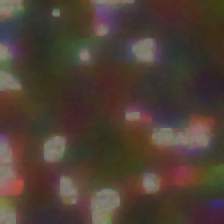}
        \caption{ConvNeXt-T\\(SSIM=0.343)}
    \end{subfigure}
    \hfill
    \begin{subfigure}{0.16\textwidth}
        \centering
        \includegraphics[height=2.5cm]{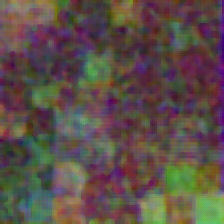}
        \caption{MaxViT-T\\(SSIM=0.214)}
    \end{subfigure}
    \hfill
    \begin{subfigure}{0.16\textwidth}
        \centering
        \includegraphics[height=2.5cm]{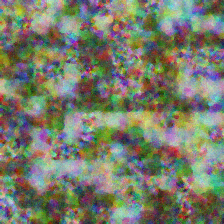}
        \caption{ViT-B/16\\(SSIM=0.017)}
    \end{subfigure}

    \caption{Gradient inversion feasibility across modern ImageNet architectures in their canonical setting (SSIM $\uparrow$).
    Reconstructions for a range of modern vision architectures, trained and evaluated on ImageNet at 224×224 resolution.
    Despite favorable conditions and extensive optimization, only Swin-T yields partial reconstructions with recognizable structure. The remaining architectures either collapse to unstructured noise or, at best, recover weak global color statistics without semantic content, illustrating that modern, performance-optimized architectures largely resist practical gradient inversion.}

    \label{fig:imagenet_architectures}
\end{figure*}

Figure~\ref{fig:imagenet_architectures} shows that gradient inversion largely fails to recover meaningful semantic information. Swin-T yields the strongest reconstruction, with partial recovery of coarse structure and color distribution. ConvNeXt-T and SwinV2-T reveal only weak traces of global color without discernible object-level structure, while MaxViT-T and ViT-B/16 collapse almost entirely to noise. SSIM separates these outcomes only coarsely, since it also rewards matching global color and low-frequency content; as described in Appendix~\ref{app:attack_robustness}, we therefore judge semantic leakage jointly with visual inspection.

Vulnerability is thus not uniform across architectures trained on the same data under comparable conditions. The contrast between Swin-T and its successor SwinV2-T is particularly informative: despite similar parameter counts and training objectives, SwinV2-T resists where Swin-T partially leaks. Among the changes SwinV2 introduces is a move from pre-normalization to residual post-normalization~\cite{liu2022swinv2}, mirroring the pre- versus post-activation contrast of Section~\ref{sec:validation}, although its other changes (such as scaled cosine attention) prevent attributing the difference to normalization alone. That the same SwinV2-T is fully vulnerable with a CIFAR-adapted stem (Section~\ref{sec:measure}) further shows that such factors act jointly rather than in isolation.

\begin{figure}[t]
  \centering
  \begin{minipage}{0.49\linewidth}
    \centering
    \includegraphics[width=\linewidth]{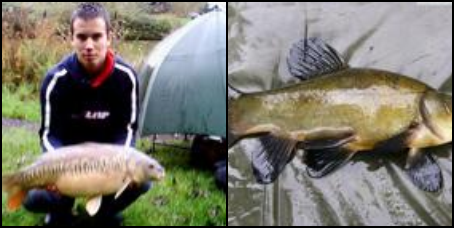}
    \vspace{-1.5em}
    \caption*{\small (a) Original \\ .}
  \end{minipage}\hfill
  \begin{minipage}{0.49\linewidth}
    \centering
    \includegraphics[width=\linewidth]{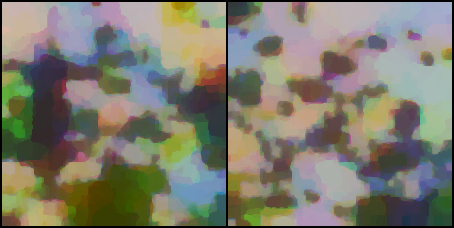}
    \vspace{-1.5em}
    \caption*{\small (b) Swin-T \\ (SSIM=0.1079)}
  \end{minipage}

  \vspace{-0.3em}
  \caption{\textbf{Gradient inversion on Swin-T with batch size two (SSIM$ \uparrow$).} Increasing the batch size from one to two is sufficient to cause reconstruction quality to collapse almost entirely to noise, illustrating the sensitivity of gradient inversion attacks to even minimal increases in inversion complexity.}
  \label{fig:swint_2img}
  \vspace{-1.0em}
\end{figure}

Feasibility is also highly sensitive to the attacker's favorable assumptions. Increasing the batch size from one to two is sufficient for Swin-T, the most vulnerable ImageNet model at batch size one, to degrade almost entirely to noise (Fig.~\ref{fig:swint_2img}), underscoring the combinatorial ambiguity introduced by even a second sample.

\noindent\textbf{Key findings.}
\begin{itemize}
    \item In their canonical configurations, modern ImageNet architectures largely resist gradient inversion even at batch size one; only Swin-T shows partial leakage, and it collapses at batch size two.
    \item Feasibility depends on architectural details such as normalization placement, and on their interaction with the stem and input resolution, so results obtained on adapted or low-resolution setups should not be extrapolated to canonical large-scale systems.
\end{itemize}

\section{Discussion}
\label{sec:discussion}

Table~\ref{tab:gia_summary} summarizes the conditions under which gradient inversion becomes feasible across our experiments.

\subsection{Architectural factors governing gradient leakage}

No single design choice determines vulnerability to gradient inversion; feasibility emerges from the interaction between normalization, architectural structure, and training setup.

Normalization plays a central role, and our results suggest a unifying reading: normalization that depends on the current input hinders inversion, while affine normalization enables it. Training-mode BatchNorm computes its statistics from the client's own input, coupling the backward pass to the quantity the attacker is trying to recover; switching to inference mode, where BatchNorm becomes affine, makes post-activation ResNet18 and YOLOv8-nano reconstructable (Sections~\ref{sec:validation}, \ref{sec:objectdetect}). Input dependence is not the only factor. Placement modulates it: pre-activation ResNet18 remains vulnerable in training mode while its post-activation counterpart resists, a contrast mirrored by pre-norm Swin-T and post-norm SwinV2-T at ImageNet scale (Section~\ref{sec:evalimagenet}). And affine normalization is not sufficient on its own, as RetinaNet resists with frozen BatchNorm. We emphasize that these are patterns across the evaluated models rather than causal claims.

Stem design and resolution alignment form a second factor. Adapting ImageNet stems to CIFAR-10 resolution makes Swin-T and SwinV2-T fully reconstructable (Section~\ref{sec:measure}), whereas SwinV2-T resists in its canonical configuration. ViT-B/16, MaxViT-T, and ConvNeXt-T resist in both, so a mismatched stem enables inversion only in combination with other factors.

Across settings, canonical, performance-optimized configurations resist: ViT-B/16, MaxViT-T, ConvNeXt-T, and SwinV2-T at ImageNet scale, and all three detectors under training-mode updates. The exceptions are pre-activation ResNet18, Swin-T at batch size one, and configurations that depart from how these models are trained in practice. Feasibility thus tracks architectural and procedural details rather than model age, and performance-driven design choices appear, as a side effect, to reduce gradient-level leakage.

\begin{table*}[t]
\centering
\small
\setlength{\tabcolsep}{4pt}
\renewcommand{\arraystretch}{1.15}
\begin{tabular}{@{}p{2.3cm} p{3.9cm} p{3.0cm} c p{3.8cm} c@{}}
\toprule
\textbf{Regime} & \textbf{Models} & \textbf{Configuration} & \textbf{Mode} & \textbf{Key factor} & \textbf{Feasible} \\
\midrule
\multirow{6}{=}{\textbf{Modern classifiers}}
& Swin-T & Canonical, ImageNet & Training & Pre-norm LayerNorm & \yes \\
& Swin-T & Canonical, ImageNet, $B{=}2$ & Training & Second sample in batch & \no \\
& SwinV2-T & Canonical, ImageNet & Training & Post-norm LayerNorm & \no \\
& ViT-B/16, MaxViT-T, ConvNeXt-T & Canonical, ImageNet & Training & -- & \no \\
& Swin-T, SwinV2-T & Adapted, CIFAR-10 & Training & Reduced stem downsampling & \yes \\
& ViT-B/16, MaxViT-T, ConvNeXt-T & Adapted, CIFAR-10 & Training & Resist despite adapted stem & \no \\
\midrule
\multirow{3}{=}{\textbf{Legacy CNNs}}
& ResNet-18 (pre-act) & Canonical & Training & Pre-activation BN & \yes \\
& ResNet-18 (post-act) & Canonical & Training & Input-dependent BN & \no \\
& ResNet-18 (post-act) & Canonical & Inference & Affine BN & \yes \\
\midrule
\multirow{4}{=}{\textbf{Object detectors}}
& YOLOv8-nano & COCO & Training & Input-dependent BN & \no \\
& YOLOv8-nano & COCO & Inference & Affine BN & \yes \\
& RetinaNet & COCO & Both & Resists even with affine BN & \no \\
& DETR & COCO & Both (identical) & LayerNorm, frozen backbone BN & \no \\
\bottomrule
\end{tabular}
\caption{\textbf{Gradient inversion feasibility across architectural and procedural regimes.} \yes~indicates meaningful reconstruction; \no~indicates failure across all evaluated settings, repeated runs, and hyperparameter search (Appendix~\ref{app:attack_robustness}). \emph{Key factor} gives the design or procedural choice distinguishing a row from its neighbours. Feasibility requires a legacy design, a resolution-adapted stem, or inference-mode client updates; among canonical modern models under training-mode updates, only Swin-T at batch size one shows partial leakage.}
\label{tab:gia_summary}
\end{table*}

\subsection{Negative results as evidence of information insufficiency}

Failures to reconstruct are systematic rather than stochastic: they persist across repeated runs, random initializations, and extensive hyperparameter search (Appendix~\ref{app:attack_robustness}), so they cannot be attributed to an unlucky optimization trajectory. Our experiments further point to two distinct mechanisms. In the first, the objective offers no target-specific direction: for all detectors in training mode, RetinaNet and DETR in inference mode, and MaxViT-T, random steps compete with or beat the step toward the target along most of the path, so an optimizer starting from noise has no signal to follow. This is a property of the objective, not of the optimizer. In the second, the objective points toward the target locally, yet the attack still fails: for ViT-B/16 and ConvNeXt-T the probe finds target-specific descent, while controlled initialization shows reconstruction quality tracking the starting point rather than the gradients. This is consistent with an underdetermined objective in which many inputs match the gradient comparably well (Section~\ref{subsec:stagnation}). In both cases, the gradients do not expose the client's input in a form that gradient matching can extract, and our negative results serve as empirical evidence of infeasibility for the configurations evaluated here.

\subsection{Why gradient inversion has stagnated algorithmically}
\label{subsec:stagnation}

Despite seven years of sustained research, the core inversion objective, matching client gradients via cosine similarity or closely related metrics, has remained essentially unchanged~\cite{carletti2025sok}.

We interpret this stagnation not as a lack of research effort, but as a reflection of the intrinsic constraints of the problem. Gradient inversion is fundamentally limited by the information content of the gradients themselves. As models grow deeper, incorporate complex mixing operations, and process increasingly rich client updates (e.g., higher resolution images or larger local batches), the combinatorial ambiguity of possible inputs explodes. In such regimes, the optimization objective becomes underdetermined: many candidate inputs yield similar gradient signals, and no amount of optimization can recover information that is no longer retrievable.

Consequently, advances in attack performance have largely been driven by relaxing assumptions such as operating in inference mode, simplifying architectures, working with low resolution images, detectable model-tampering, or introducing additional side information such as pretrained image priors~\cite{hatamizadeh2022gradvit}, rather than by algorithmic breakthroughs that extract fundamentally new information from gradients. Our results suggest that without new leakage channels, substantial leaps in GIA feasibility are unlikely.

\subsection{Limitations}
\label{subsec:limitations}

Apart from the ResNet18 validation of Section IV, which uses batches of 16 CIFAR-10 images, our evaluation uses batch size one and a single local step, the most favorable FedSGD setting for the attacker, and assumes full label knowledge; the exact-statistic inversion of Eq.~\ref{eq:attack_exact} likewise holds for a single step at known momentum. Targets are selected to be vulnerable rather than average-case, so our results bound rather than estimate typical leakage. YOLOv8-nano is evaluated at random initialization. Finally, the probe is a one-sided screen, and its competitors are sampled at a fixed fraction of the input resolution: at $224\times224$ they are too weak to compete with the target step for any model ($\nu = 0$ throughout, including resistant models), so the probe does not discriminate at ImageNet scale. Competitors whose dimensionality does not grow with resolution, or the optimizer's own descent direction, are a natural next step.

Each setting is evaluated on a single target, and the ImageNet target was selected on Swin-T, which may favor its vulnerabilities over those of other architectures; distributions over many targets selected per architecture would strengthen these conclusions. Success is judged partly by visual inspection, since SSIM separates regimes only weakly. Stem and resolution effects are entangled in our CIFAR-10 versus ImageNet comparison, and the architectural comparisons of Section~\ref{sec:discussion} differ in several respects at once, so controlled ablations would be needed to isolate individual factors. We use recovered normalization statistics as a soft regularizer; stronger uses, such as hard constraints, remain unexplored. Finally, common medical-imaging setups (small CNNs, low resolution, pre-activation designs) resemble configurations we found vulnerable, so our conclusions should not be read as covering them.

Our threat model also gives the server individual client updates; secure aggregation, which exposes only the sum over clients, and differential privacy, which perturbs updates, would further constrain the attacks we evaluate, so our results describe the unprotected worst case.

\section{Conclusion and Future Work}
\label{sec:conclusion}

We presented a feasibility analysis of optimization-based gradient inversion under an honest-but-curious server. To make negative results trustworthy, we strengthened the training-mode attack by recovering the client's exact normalization statistics, and introduced the target-interpolation feasibility probe, a lightweight diagnostic that tests whether the attack objective singles out the direction toward the target, grading resistance that a full attack reports only as failure. Using both, we conducted, to our knowledge, the first evaluation of gradient inversion on object detection.

Across classification and detection, feasibility is governed by a small set of interacting factors: normalization type and placement, stem design and resolution alignment, and train- versus inference-mode operation. No single factor is decisive, so gradient inversion risk is best assessed case by case, but the overall tendency is clear: canonical, performance-optimized configurations resist. Where attacks fail, our results point to two mechanisms, both forms of information insufficiency: an objective that offers no target-specific direction, and one that points toward the target locally but remains underdetermined, so that even a highly informative initialization yields a reconstruction only as good as the initialization itself.

Taken together, and consistent with an emerging consensus in the field, our results indicate that, for image classification and detection, image reconstruction via optimization-based gradient inversion without learned priors is improbable in modern, performance-optimized configurations. The successes we observe require legacy designs, resolution-adapted stems, inference-mode client updates, or, for Swin-T, a batch of a single image; adding a second image already suffices to defeat the most vulnerable canonical ImageNet model we evaluated. Demonstrating attack success under such conditions should not be conflated with practical privacy risk. This matters beyond the research community: as the case that motivated this work illustrates, decisions to adopt or abandon federated learning in sensitive domains rest on exactly such risk assessments.

Our results do not imply that gradients are free of privacy-relevant information, only that the semantic leakage targeted by classical reconstruction attacks becomes increasingly constrained as system complexity grows. We see three directions for future work:
\begin{itemize}
    \item \emph{A calibrated robustness score.} Wherever the probe flags an objective as non-specific, full attacks fail, and on the detectors its graded output follows the attack outcomes, at a fraction of their cost. Competitors whose dimensionality does not grow with input resolution, the optimizer's own descent direction, and decoy endpoints that test for underdetermination could extend it into a reliable, attack-free robustness score that practitioners and regulators can compute before deployment.
    \item \emph{Realistic priors.} Replacing the noise-corrupted target of our controlled-initialization study with priors of graded realism, such as generative samples or nearest neighbours from auxiliary data, would measure how closely practical prior-based attacks approach our upper bound.
    \item \emph{Subtler leakage.} Identifying what remains extractable once full reconstruction is no longer feasible, such as attribute signals, distributional properties, or cross-round correlations.
\end{itemize}


\section*{Open Science}
The implementation of our Gradient Inversion Attack examples are publicly available at: \url{https://github.com/aidotse/LeakPro}

\section*{LLM usage considerations}
The authors used generative AI-based tools to revise the text, improve flow, and correct typos, grammatical errors, and awkward phrasing. LLMs were used for editorial purposes in this manuscript, and all outputs were inspected by the authors to ensure accuracy and originality.

\bibliographystyle{IEEEtran}
\bibliography{references}

\appendix
\section{Attack Robustness and Reproducibility}
\label{app:attack_robustness}

Given the prevalence of negative results in our evaluation, it is essential to establish that observed failures of gradient inversion are not artifacts of poor initialization, suboptimal hyperparameter choices, or insufficient optimization effort. This appendix summarizes the measures we take to ensure that our attack evaluations are robust, reproducible, and representative of best-practice gradient inversion methodology.

Across all experimental settings, we perform multiple independent attack runs with different random initializations and reconstruction seeds. For each setting, we further explore a broad range of attack hyperparameters commonly used in the literature, including regularization weights and optimizer configurations. Rather than fixing a single configuration, we actively search for settings that maximize reconstruction quality, measured primarily using SSIM~\cite{wang2004image}, which is widely used as a perceptual similarity metric and correlates well with human visual assessment, and secondarily by visual inspection of semantic structure. This ensures that failures cannot be attributed to poor initialization or attack parameter choices.

Hyperparameter tuning is performed using Bayesian optimization via the Optuna framework~\cite{akiba2019optuna}, employing its default Tree-structured Parzen Estimator (TPE) sampler. This allows efficient exploration of large, continuous parameter spaces and prioritizes promising regions based on past evaluations.

To account for variability across client data, we do not evaluate attacks on a single fixed example. Instead, we first identify a challenging target by sampling multiple candidate inputs and selecting the example that yields the strongest reconstruction under a strong reference architecture. This selection is performed jointly with hyperparameter tuning using Bayesian optimization, and reconstruction quality is measured using SSIM. The selected input is then used consistently across architectures to ensure that observed differences reflect architectural effects rather than differences in data difficulty. This procedure reflects a realistic adversary seeking to identify vulnerable samples rather than average-case behavior.

\subsection{Concrete example (Figure~\ref{fig:imagenet_architectures}).}

For the ImageNet experiment shown in Figure~\ref{fig:imagenet_architectures}, we proceed as follows. We first randomly sample five candidate images from the dataset and use Swin-T as a reference model to identify a vulnerable data point among these candidates.

The choice of Swin-T as the reference architecture is motivated by extensive preliminary experiments across all considered models, in which we observed that Swin-T consistently exhibited the strongest reconstruction performance and highest susceptibility to gradient inversion. These initial studies involved running numerous attack trials across models and randomly sampled data points, and indicated that Swin-T represents a near worst-case architecture from an attacker’s perspective among the selected models. We therefore adopt Swin-T as a conservative reference model for selecting challenging and vulnerable samples.

For each candidate image, we optimize over key attack hyperparameters, including:
(i) the total-variation regularization weight $\lambda_{TV} \in [10^{-6}, 10^{-1}]$ (log-uniform),
(ii) normalization statistic regularization strength $\lambda_{BN} \in \{0, 10^{-4}, 10^{-2}\}$,
(iii) attack learning rate $\eta \in [10^{-4}, 10^{2}]$ (log-uniform), and (iv) the data point index.

Hyperparameter exploration is performed using Bayesian optimization with multiple trials, and for each configuration we perform multiple independent attack runs with different random initializations. The candidate yielding the strongest reconstruction based primarily on SSIM and confirmed by visual inspection to ensure that high scores correspond to meaningful semantic leakage rather than trivial artifacts is selected as the representative example.

This selected image is then used consistently across \emph{all} models in Figure~\ref{fig:imagenet_architectures}. For each architecture, we perform 30 independent attack runs with different random initializations, again tuning the same attack hyperparameters (excluding the data point index, which is fixed). This ensures that the reported cross-model differences reflect architectural effects rather than differences in data difficulty or attack configuration.

Reconstruction quality is assessed quantitatively using SSIM and qualitatively through visual inspection. Visual inspection is used to distinguish meaningful visual leakage such as reconstructed patches or semantic structure from trivial artifacts such as global color matching or noise patterns, which may achieve moderate SSIM scores without revealing interpretable content. The reconstruction presented in each figure is therefore selected based on both quantitative and qualitative criteria.

This methodology ensures that when attacks fail, such failures are consistent across repeated runs and not sensitive to a particular random seed, initialization, or hyperparameter configuration. In settings where gradient inversion is genuinely feasible, we observe that semantic structure typically emerges early and consistently during optimization. For example, in experiments on Swin-T, recognizable structure appears within the first few optimization iterations and remains stable across runs. This behavior provides a strong early indicator of attack success and suggests that extensive hyperparameter exploration or large numbers of trials are not required to detect reconstructability when meaningful leakage is present. Consequently, the additional trials reported in this work should be interpreted as a conservative procedure to rule out optimization artifacts, rather than a prerequisite for observing successful reconstructions.

Together, these procedures ensure that our reported results, both positive and negative, reflect intrinsic properties of the evaluated architectures and training pipelines, rather than limitations of the attack implementation or optimization process.

\subsection{Feasibility probe protocol}
\label{app:probe}
All probe results use $n{=}10$ noise draws $x_0 \sim \mathcal{N}(0, I)$ in the normalized input space, a path of 141 points with $\alpha\in[-0.2,1.2]$, and the specificity test at every fifth point with $K{=}10$ smooth directions (Gaussian noise at one quarter resolution, bilinearly upsampled, unit norm over the whole batch). The same directions are used for every model, so draws are paired across models. The objective is the attack's own, evaluated at the probe point with the attack's client gradient and regularizer hooks, and checked to agree with the attack's loss to floating-point precision; the cosine term is computed in double precision. For the detectors we use the attacked image with all of its oracle boxes. For training mode, RetinaNet's frozen BatchNorm layers are replaced by trainable BatchNorm with the same parameters. DETR's backbone BatchNorm is frozen and dropout is disabled throughout, so its training mode coincides with inference mode.

\end{document}